\documentclass[12pt]{JHEP3}

\usepackage{epsfig}
\epsfclipon

\def\Dsl{\hbox{/\kern-.6700em\it D}} 
\def\dsl{\hbox{/\kern-.5300em$\partial$}}
\def\ol#1{\overline{#1}}

\def\sla#1{\hbox{/\kern-.6700em #1}}

\def\eq{\begin{equation}}
\def\eeq{\end{equation}}
\def\eqa{\begin{eqnarray}}
\def\eeqa{\end{eqnarray}}

\def\bd{\begin{displaymath}}
\def\ed{\end{diplaymath}}

\def\Box{ {\,\lower 0.9pt\vbox{\hrule\hbox{\vrule height0.2cm \hskip 0.2cm \vrule height 0.2cm }\hrule}\,}}
\def\lsim{{\ \lower-1.2pt\vbox{\hbox{\rlap{$<$}\lower5pt\vbox{\hbox{$\sim$}}}}\ }}
\def\gsim{{\ \lower-1.2pt\vbox{\hbox{\rlap{$>$}\lower5pt\vbox{\hbox{$\sim$}}}}\ }}

\def\pref#1{(\ref{#1})}

\def\ol#1{\overline{#1}}
\def\Dsl{\hbox{/\kern-.6700em\it D}} 
\def\dsl{\hbox{/\kern-.5300em$\partial$}}

\def\ie{{\it i.e.}}

\def\beginvector{\left( \begin{array}{c} }
\def\endvector{\end{array} \right)}
\def\endignore{}
\def\ignore#1\endignore{}

\def\Sph2{{\mathcal S}^2}

\def\ie{{\it i.e.}}

\title{Phenomenological Constraints on Extra-Dimensional Scalars}
\author{G. Azuelos,${}^{1,2}$ P.-H. Beauchemin${}^{1,3}$ and
C.P.~Burgess${}^3$
\\

${}^1$ Laboratoire Ren\'e J.-A. L\'evesque,
 Universit\'e de Montr\'eal, \\
 C.P. 6128, succ. centre-ville, Montr\'eal, Qu\'ebec, Canada, H3C
 3J7.\\

${}^2$ Triumf,
4004 Wesbrook Mall, Vancouver BC, V6T 2A3.\\

${}^3$ Physics Department, McGill University,
                3600 University Street,\\
                Montr\'eal, Qu\'ebec, Canada, H3A 2T8.}

\abstract{We examine whether the ATLAS detector has sensitivity 
to extra-dimensional scalars (as opposed to components of
higher-dimensional tensors which look like 4D scalars), in
scenarios having the extra-dimensional Planck scale in the TeV
range and $n \ge 2$ nonwarped extra dimensions. Such scalars appear
as partners of the graviton in virtually all higher-dimensional
supersymmetric theories. Using the scalar's lowest-dimensional
effective couplings to quarks and gluons, we compute the rate for
the production of a hard jet together with missing energy. We find
a nontrivial range of graviscalar couplings to which ATLAS could
be sensitive, with experiments being more sensitive to couplings
to gluons than to quarks. Graviscalar emission increases the
missing-energy signal by adding to graviton production, and so
complicates the inference of the extra-dimensional Planck scale
from an observed rate. Because graviscalar differential cross
sections resemble those for gravitons, it is unlikely that these
can be experimentally distinguished from one another should a
missing energy signal be observed.}

\preprint{McGill-03/17}

\begin{document}

\section{Introduction}
The realization that gravitational physics could be as low as the
TeV scale \cite{LowStringScale,ADD,RS}, has sparked considerable
interest in the possibility of detecting extra-dimensional
gravitons in upcoming accelerator experiments
\cite{realgraviton,virtualgraviton,OpPollution}. There are now
several types of models within this class, which largely differ in
the masses which they predict for extra-dimensional states. In the
extreme ADD scenario \cite{ADD} these Kaluza-Klein masses can be
as small as $m_{KK} \sim 10^{-3}$ eV, while in the warped RS
picture \cite{RS,warped6D} they are as large as the TeV scale.

Although the graviton has grabbed most of the phenomenological
attention, models with low string scales often also predict many
other kinds of extra-dimensional physics which might also be
detectable. For instance, many of the best-motivated models are
inspired by string theory, and so often predict the gravitational
physics of the bulk to be supersymmetric and thus contain entire
extra-dimensional gravity supermultiplets. Indeed the
supersymmetry of the gravitational sectors of these models may be
one of their great strengths, perhaps helping to control the
contributions of quantum corrections to the effective 4D
cosmological constant \cite{branesphere}.

Since the relevant supersymmetry is extra-dimensional, from the
four-dimensional perspective the graviton supermultiplet forms a
representation of {\it extended} supergravity (\ie\ supergravity
with more than one supersymmetry generator). As such, the graviton
multiplet in these theories can contain several spin-3/2
gravitini, spin-1 gauge bosons, spin-1/2 fermions and scalars, in
addition to the graviton. For example a typical gravity
supermultiplet in 6 dimensions already contains 4D particles
having all spins from 0 through 2. Although much less is known
about the phenomenological consequences of these other modes ---
see, however, \cite{SusyLED} --- generically they may be expected
to have masses and couplings which are similar to the graviton's
since they are related to it by extra-dimensional supersymmetry.

In this paper we focus on the properties of the graviscalar, which
we take to mean an honest-to-God extra-dimensional scalar which is
related to the graviton by supersymmetry. (We do {\it not}, for
example, mean a 4D scalar which arises as an extra-dimensional
component of the metric tensor itself --- although many of our
results will also apply to such a particle \cite{GRW}.) Our goal
is to identify the relevant couplings of such a scalar, and to use
these to explore its experimental implications. In particular, we
identify how the production of such a scalar can compete with the
predictions for graviton production.

For phenomenological purposes we concentrate on the implications
of real graviscalar production, rather than calculating the
consequences of its virtual exchange. We do so for the same
reasons as also apply to virtual exchanges of gravitons
\cite{virtualgraviton} and graviphotons \cite{SusyLED}: their
virtual exchange cannot be distinguished from the effects of local
interactions produced by other kinds of high-energy physics (such
as the exchange of massive string modes) \cite{OpPollution}.

We organize the presentation of our results as follows. In the
next section we describe the most general possible low-energy
couplings of an extra-dimensional scalar with quarks and gluons,
and summarize the domain of validity of an effective-field-theory
analysis. Section 3 then uses these couplings to compute the
relevant cross sections at the parton level, and for jet plus
missing energy production in proton-proton collisions. Section 4
summarizes the results of numerical simulations based on the cross
section of section 3. It compares the predicted production rate
both with the expected Standard Model backgrounds, and with
previously-calculated graviton emission rates. Our conclusions are
then briefly summarized in section 5.

\section{Low-Energy Graviscalar Couplings}
The class of models of present interest are those for which all
Standard-Model particles reside on a 4-dimensional surface (or
3-brane) sitting in a higher-dimensional space. We would like to
know how such particles can couple to higher-dimensional scalars
which are related to the graviton by supersymmetry. In particular,
since we look for direct experimental signatures in accelerators,
we concentrate on trilinear interactions involving a single
higher-dimensional scalar and two Standard-Model particles.

At low energies we may parameterize the most general such
couplings by writing down the most general lowest-dimension
interactions which are consistent with the assumed low-energy
particle content and symmetries of the theory. 
In this paper, we are concerned with the production of scalars in parton 
interactions.
The most general couplings to quarks and gluons are given by the following
Lagrangian:
\begin{eqnarray} \label{E:efflagrangian}
   \lefteqn{\mathcal{L}_{\mathtt{EFF}}  =  \partial_M\phi(x,y) \,
   \partial^M\phi(x,y) } \nonumber \\
   & & \qquad\qquad  -\delta^n(y)  \, \left[
     \sum_{Q} \bar{\Psi}_Q^i(x)(\overline{g} +i\overline{g}_5\gamma_5)_{ij}
     \Psi_Q^j(x)\phi(x) \right.
       \\
   & & \qquad\qquad \left.  + ~
     \overline{c}_g \, G_{a}^{\mu \nu}(x)G^{a}_{\mu \nu}(x)\phi(x)
    + \overline{b}_g  \epsilon^{\mu\nu\lambda\rho} {G}^{a}_{\mu \nu}(x)G^{a}_{\lambda \rho}(x)
    \phi(x) \right.
     \nonumber \\
    & & \qquad\qquad  + ~\overline{c}_{\gamma} F^{\mu \nu}(x)F_{\mu \nu}(x) \phi(x) +
    \overline{b}_{\gamma}  \epsilon^{\mu\nu\lambda\rho}
    {F}_{\mu \nu}(x)F_{\lambda \rho}(x) \phi(x)
  \Bigg] \, , \nonumber
\label{equ:lagrangian}
\end{eqnarray}

with arbitrary dimensionful coupling parameters
$\overline{g}_{ij}$, $(\overline{g}_5)_{ij}$, $\overline{c}_g$,
$\overline{b}_g$, $\overline{c}_\gamma$ and $\overline{b}_\gamma$.
The indices $i,j = 1,2,3$ here label the Standard Model's three
generations.

In what follows it is convenient to define dimensionless couplings  by
scaling out the appropriate power of the reduced Planck mass in  D
dimensions,\footnote{More precisely, $\overline{M}_D$ is the  reduced
Planck mass in $D=4+n$ dimensions, defined in terms of the
$D$-dimensional Newton's constant by $8 \pi \, G_D =
\overline{M}_D^{2-D}$.} according to  ${g}_{ij} = \overline{g}_{ij}
\overline{M}_D^{n/2}$,  $c_g = \overline{c}_g \overline{M}_D^{1+n/2}$
and so on.  It should be kept in mind when doing so, however, that the
fermion  interactions are not $SU_L(2) \times U_Y(1)$ invariant and so
the size to  be expected for the couplings $\overline{g}_{ij}$ and
$(\overline{g}_5)_{ij}$ depends strongly on the way in which the
electroweak gauge group is broken in the underlying theory which
produces  them. In particular, if the new physics is not involved in
electroweak  symmetry breaking then  there is a natural suppression of
the fermionic dimensionless  couplings, $g_{ij} \sim
v/\overline{M}_D$, where $v = 246$ GeV.  (If the relevant
dimensionless couplings of the underlying model are  similar in size
to Standard Model yukawa couplings, then this  suppression can be even
smaller, although they need not be  this small in all models.)  In
principle, if the extra-dimensional physics  were itself the sector
which broke the electroweak gauge group, even  the suppression by
powers of $v/\overline{M}_D$ might not be present,  although  in this
case the new-physics scale cannot be very large  compared to $v$

In these expressions, the coordinates $x^\mu$ describe
the 4 dimensions parallel to the Standard-Model brane and
$y^m$ similarly describes the $n$ transverse dimensions. The brane
itself is assumed to be located at the position $y^m = 0$ in these
extra dimensions. $F_{\mu\nu}$ denotes the usual electromagnetic
field strength, and $G_{\mu\nu}^a$ is the non-abelian gluon field
strength, $G^{a}_{\mu \nu} = \partial_{\mu}G^{a}_{\nu} -
\partial_{\nu}G^{a}_{\mu} + g_3 f^{abc}G_{\mu}^bG_{\nu}^c$, with
$g_3$ denoting the QCD coupling constant. Finally, $\Psi_Q$
generically denotes any of the Standard-Model fermion mass
eigenstates, whose electric charge is denoted by $Q$.

Several comments concerning these effective interactions bear
emphasis:

\medskip\noindent$\bullet$
As written, the fermion interaction of eq.~\pref{E:efflagrangian}
appears not to be invariant under electroweak gauge
transformations. This is because we have already replaced an
explicit factor of the Standard Model Higgs field with its vacuum
expectation value, $v = 246$ GeV, and rotated to a fermion mass
eigen-basis. Although the resulting couplings, $g_{ij}$ and
$(g_5)_{ij}$, can in principle be off-diagonal and so involve
flavour-changing neutral currents, we will assume these not to
arise since they would be strongly constrained if present.

\medskip\noindent$\bullet$
  Keeping in mind that in real models the scalar in
question typically lies in a supermultiplet with the graviton and
so couples with similar strength, the interactions written above can
be regarded as of leading order in inverse powers of $\overline{M}_D$, and
are expected to break down once physical energies approach this scale.
One exception would be the possibility of a direct
mixing term between the extra-dimensional scalar and the Standard
Model Higgs. Since the following analysis concentrates on the
phenomenology of jets plus the graviscalar, we ignore this kind of
$\phi$-Higgs mixing in what follows. For similar reasons we also
ignore graviscalar couplings to the $Z$ and $W$ bosons.

\medskip\noindent$\bullet$
%
   Derivative couplings to the fermions are not included here since such
terms are not independent, as they can be re-written in the form
given above  by performing a field redefinition and are
expected to be small~\cite{PhysRep}.

\medskip\noindent$\bullet$
The factor $\delta^n(y)$ in eq.~\pref{equ:lagrangian} expresses
explicitly the broken translation invariance of the bulk due to
the presence of the brane. Consequently momentum transverse to the
brane is not conserved, allowing bulk particles to be emitted into
the extra dimensions even if all of the initial particles of the
interaction were themselves confined to the brane. Physically, the
unbalanced transverse momentum is absorbed by the recoil of the
brane itself, with no energy cost because of the brane's enormous
mass.

\medskip\noindent$\bullet$
  Our effective Lagrangian, as written in eq.~\pref{E:efflagrangian},
applies equally well to both warped~\cite{RS} and unwarped~\cite{ADD} 
models. The spectrum of effective 4D masses
emerges from the expansion of the bulk scalar field
$\phi$ in terms of a basis of Kaluza-Klein (KK) modes: 
$\phi(x,y) = \sum_k \, \phi_k(x) \, u_k(y)$.
Below, we assume large extra dimensions, allowing us
to integrate over the phase space of all bulk particles using a flat
geometry (for a more explicit
computation of the limitations of this approximation, see
ref.~\cite{fred}.) 
\section{Physical Predictions}
The effective Lagrangian of eq.~\pref{E:efflagrangian} can be used
to compute the cross-section at tree-level for production of
graviscalars in $p$-$p$ collisions at the Large Hadron Collider
(LHC). We shall restrict our attention to hadronic production
mechanisms since these are expected to dominate at LHC energies,
and so might be expected to give the clearest signal. For these
purposes our interest at the parton level is therefore in the
reactions $\bar{q} q \to \phi X$, $gg \to \phi X$ and $qg \to \phi
X$, where $X$ represents a Standard-Model final state such as a
single well-defined jet. The physical signal which such a reaction
would produce is a well defined jet plus missing energy as the
graviscalar escapes into the extra dimensions. 

\subsection{Parton-Level Cross-Sections}
Using these Feynman rules (given in Appendix), we next compute the parton-level cross
section for producing a hard quark or gluon plus a graviscalar in
the final state. There are three processes which are relevant: $qq
\rightarrow g \phi$, $qg \rightarrow q \phi$ and $gg \rightarrow
g\phi$.

It is convenient to divide the higher-dimensional graviscalar
momentum vector, $\ell_{4+n}^M$ with $M =0,...,3+n$, into its
continuous 4-dimensional components, $\ell_4^\mu$ with $\mu =
0,...,3$, and its quantized $n$-dimensional components, $L^m$ with
$m = 4,...,3+n$. The total squared-momentum for a massless bulk
graviscalar is then $\ell_{4+n}^2 = \ell_4^2 + L^2=0$, which we
write as $\ell_4^2 =-M^2$, where $M$ is the effective
4-dimensional mass due to the particle's motion in the extra
dimensions. As mentioned earlier, we perform our calculations in
the approximation where the length scales associated with the
extra dimensions are much larger than the wavelengths of the
partons involved, since this is a good approximation for the
applications of interest and allows a tractable treatment of the
graviscalar phase space in the extra dimensions. (However, as a
consequence, our further results do not generalize to warped
models.) With the approximation the quantization of $L^m$ is not
important, and sums over this variable may be approximated as
integrals. The corresponding integration measure is then:
\begin{equation}
 \frac{d^nL}{(2\pi)^n}=  \frac{(L^2)^{({n-2})/{2}} \,
 }{2\,(2\pi)^n} \, dL^2 \, d\Omega_n \, .
\end{equation}
After integration over the angular degrees of freedom, $\Omega_n$,
the final phase-space measure used for the graviscalar bulk
momentum becomes:
\begin{equation}
  \int_{\Omega_n}\frac{d^nL}{(2\pi)^n}=\frac{(M^2)^{({n-2})/{2}} \, dM^2}
  {2\, \Gamma(\frac{n}{2}) \, (2\pi)^{{n}/{2}}} \, .
\end{equation}

Evaluating the Feynman graphs of Fig.~\pref{fig:feyngraph} (see Appendix) to
obtain the parton-level differential cross sections gives the
following results. The quark-annihilation channel is obtained by
evaluating graphs (a) through (c), which give:

\begin{eqnarray}
  \frac{d\sigma(q\bar{q}\rightarrow g \phi)}{d\hat{t} \, d\hat{u} \, dM^2} &=&
  \frac{\alpha_{s}(2\pi)^{{n}/{2}}(M^2)^{({n-2})/{2}}}{18 \,
  \Gamma(\frac{n}{2}) \, M_D^n \, \hat{s}^{2}}
  \left[\frac{(g^2 + g_{5}^{2})}{(2\pi)^{{2n}/({2+n})}} \left(
  \frac{2M^2 \hat{s} +
      (\hat{u} + \hat{t})^2}{\hat{u} \hat{t}} \right) \right.
      \\
   && \qquad\qquad\qquad\qquad\qquad \left. + \frac{4(c^2 + b^2)}
   {M_D^{2}} \left( \frac{\hat{t}^2+\hat{u}^2}{\hat{s}} \right) \right]
  \delta (\hat{s}+\hat{t}+\hat{u}+M^2) \, . \nonumber
\end{eqnarray}
Notice that helicity conservation precludes graph (c) interfering
with graphs (a) and (b) in the limit where parton masses are
neglected (as we assume).

The quark-gluon scattering contribution is similarly obtained by
evaluating the graphs (d), (e) and (f), to give
\begin{eqnarray}
  \frac{d\sigma(q g\rightarrow q \phi)}{d\hat{t} \, d\hat{u} \, dM^2} &=&
 - \, \frac{\alpha_{s}(2\pi)^{{n}/{2}}(M^2)^{({n-2})/{2}}}{48 \,
 \Gamma(\frac{n}{2}) \, M_D^n \, \hat{s}^{2}}
  \left[\frac{(g^2 + g_{5}^{2})}{(2\pi)^{{2n}/({2+n})}}
  \left( \frac{\hat{u}^2+M^4}
    {\hat{s} \hat{t}} \right) \right. \\
    &&\qquad\qquad\qquad\qquad \left.
    + \frac{4(c^2 + b^2)}{M_D^{2}} \left(
    \frac{\hat{t}^2+\hat{s}^2}{\hat{u}} \right)
    \right]\delta (\hat{s}+\hat{t}+\hat{u}+M^2) \, . \nonumber
\end{eqnarray}
As before, the neglect of parton masses ensures the
non-interference of graph (f) with graphs (d) and (e).

Evaluating the gluon fusion graphs, (g) through (j), gives
\begin{eqnarray}
  \frac{d\sigma(g g\rightarrow g \phi)}{d\hat{t} \, d\hat{u} \, dM^2} &=&
                  \frac{3\,\alpha_{s}(2\pi)^{{n}/{2}}
    (M^2)^{({n-2})/{2}}}{16 \, \Gamma(\frac{n}{2}) \, \hat{s}^{3} \,
    \hat{t} \, \hat{u}} \left( \frac
       {(c^2 + b^2)}{M_D^{n+2}} \right)
 \left[(\hat{u}+\hat{t})^4 +(\hat{u}+\hat{s})^4 +(\hat{t}+\hat{s})^4
 \phantom{\frac12} \right. \nonumber \\
 && \left. +12\, \hat{s} \,
 \hat{t} \, \hat{u} \, M^2 + \frac{1}{2}
    (\hat{t}^2\hat{u}^2 +\hat{t}^2\hat{s}^2+\hat{u}^2\hat{s}^2)\right]\delta
    (\hat{s}+\hat{t}+\hat{u}+M^2)\, .
\label{equ:cross-sect}
\end{eqnarray}
Notice that both the expressions for the gluon-fusion and
quark-annihilation processes are invariant under the exchange
$\hat{t} \leftrightarrow \hat{u}$, as is expected on general
grounds from the charge-conjugation invariance. The mass scale
$M_D$ used here is related to the scale $\ol{M}_D$ defined earlier
by $M_D^{n+2} = (2\pi)^n \ol{M}_D^{n+2}$.

Since these expressions depend only on the dimensionless coupling
combinations $g^2 + g_5^2$ and $c^2 + b^2$, in what follows we set
$g_5 = b = 0$ and choose $c \ge 0$ and $g \ge 0$ without loss of
generality.

\subsection{Proton-Proton Cross Sections}
The cross section for proton-proton collisions is obtained from
the parton-level results just calculated in the usual way, by
convoluting with the parton distribution functions, $f_i(x,Q^2)$.
For our later analysis we compute the cross section for the
process $pp \to \phi + \hbox{jet}$, which has the form:
\begin{equation}
  \sigma = \sum_{ij}
  \int  dx_1 \, dx_2 \, d\hat{t} \, dM^2 \; f_i(x_1,Q^2) \, f_j(x_2,Q^2)
  \left. \frac{d\sigma(ij \to \phi X)}{d\hat{t}\, dM^2} \right|_{\hat{s} = x_1 \,
  x_2 \, E_{cm}}  \, ,
\end{equation}
where the sum on $i,j$ runs over the types of partons available in
the proton and $X$ corresponds to an energetic quark or gluon. We
generate events randomly to which we assign weights by performing
the integration. These events are then accepted or rejected
according to their weight by the generator PYTHIA~\cite{PYTHIA}, which
also performs hadronization.

The phase-space integrals are performed subject to the following
constraints.
\begin{itemize}
\item We require the final jet's transverse momentum to satisfy
$P^2_T > P_{cut}^2$, where $P_{cut}$ ($=E_{T,jet}^{min}$), is a
minimum value which we specify below. Using the kinematical
relation $P_T^2 = {\hat{t} \, \hat{u}}/{\hat{s}}$, where $\hat{u}
= M^2 - \hat{t} - \hat{s}$, this leads to the following
integration limits for $\hat{t}$: $t_{min} < \hat{t} < t_{max}$,
where
\begin{eqnarray}
 t_{min} &=& \frac12 \Bigl[ {(M^2-\hat{s}) -\sqrt{(M^2-\hat{s})^2
 -4P_{cut}^2\hat{s}}} \, \Bigr] \\
 \hbox{and} \quad t_{max} &=& \frac12 \Bigl[ {(M^2-\hat{s}) +
 \sqrt{(M^2-\hat{s})^2-4P_{cut}^2\hat{s}}} \, \Bigr] \, .
\end{eqnarray}

\item  Energy-momentum conservation implies the following upper limit for $M^2$
\begin{equation}
            0 \leq M^2 \leq M^2_{max} = \hat{s} - 2\sqrt{\hat{s}}P_{cut} \, .
\end{equation}
\item Finally, the integration limits on the parton energy fractions are:
\begin{equation}
  x_{min} \equiv \frac{\hat{s}_{min}}{s}
  = \frac{4 P_{cut}^2} {s}
  \leq x_1,x_2 \leq 1 \, ,
\end{equation}
where as usual $s$ denotes the invariant initial energy of the
proton-proton collision.
\end{itemize}
We have checked our numerical integration by recomputing the
differential cross-section for graviton production, using the
parton-level cross sections computed by Giudice {\it et al}
\cite{realgraviton}, and comparing with their results.

\subsection{Validity of the Low-Energy Approximation}
As discussed above, our expressions for the parton-level reaction
cross sections are only valid for parton energies well below the
cut-off scale, $M_D$. But since we can only specify the initial
proton energies (and phase-space cut-offs like $P_{cut}$), we
cannot know for sure whether the numerical integration includes
parton reactions which carry energies which are too large. If so,
our calculation becomes unreliable in that part of phase space for
which the probability of very-energetic parton processes cannot be
neglected. In this section we define a proton-level criterion for
estimating the extent to which high-energy parton processes
pollute our calculations in various parts of phase space. Our goal
in so doing is to be able to choose upper limits for quantities
like $P_{cut}$ which minimize this pollution.

\begin{figure}[htbp]
  \begin{center}
    \mbox{\epsfig{file=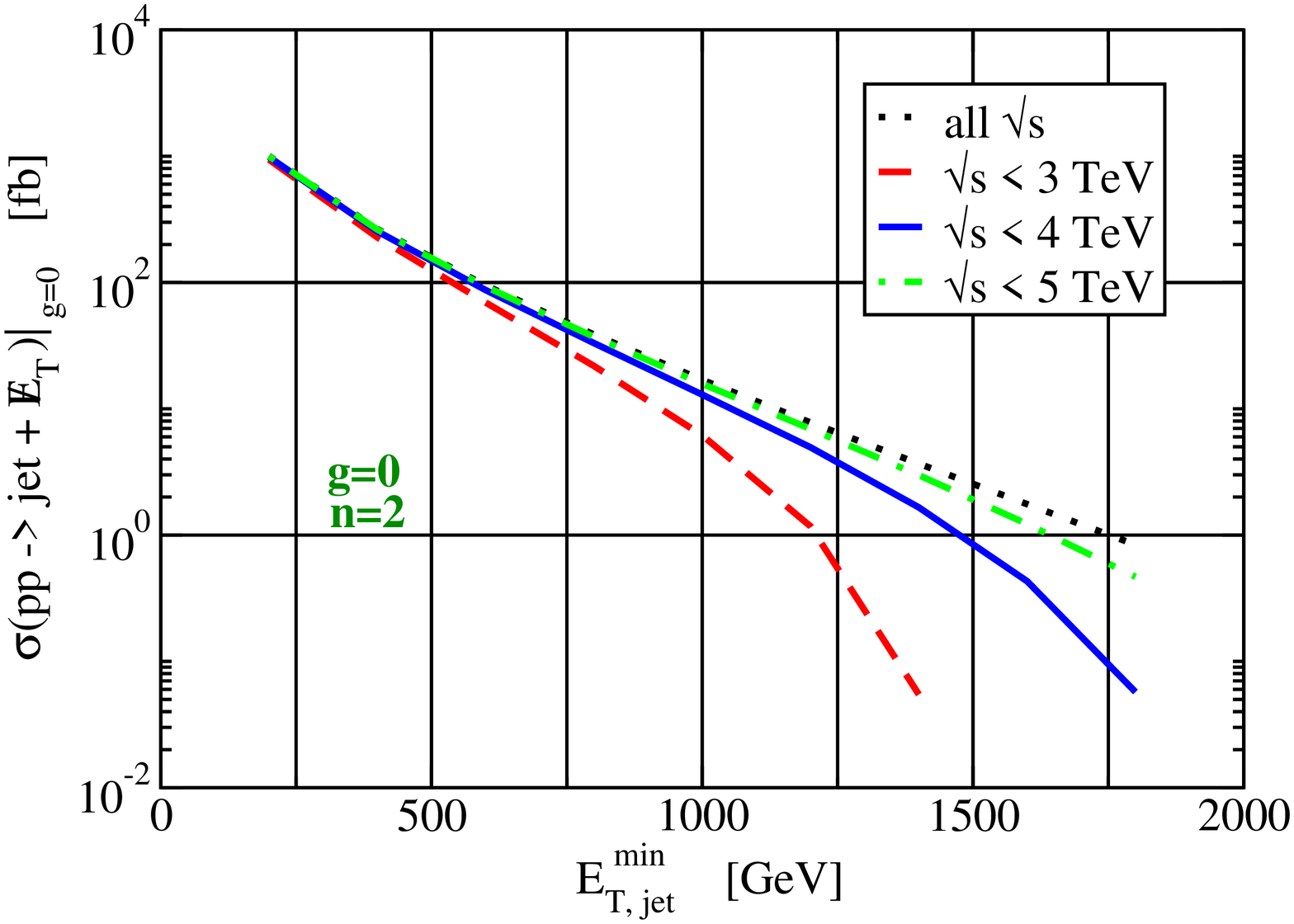, height=6.0cm}}
    \mbox{\epsfig{file=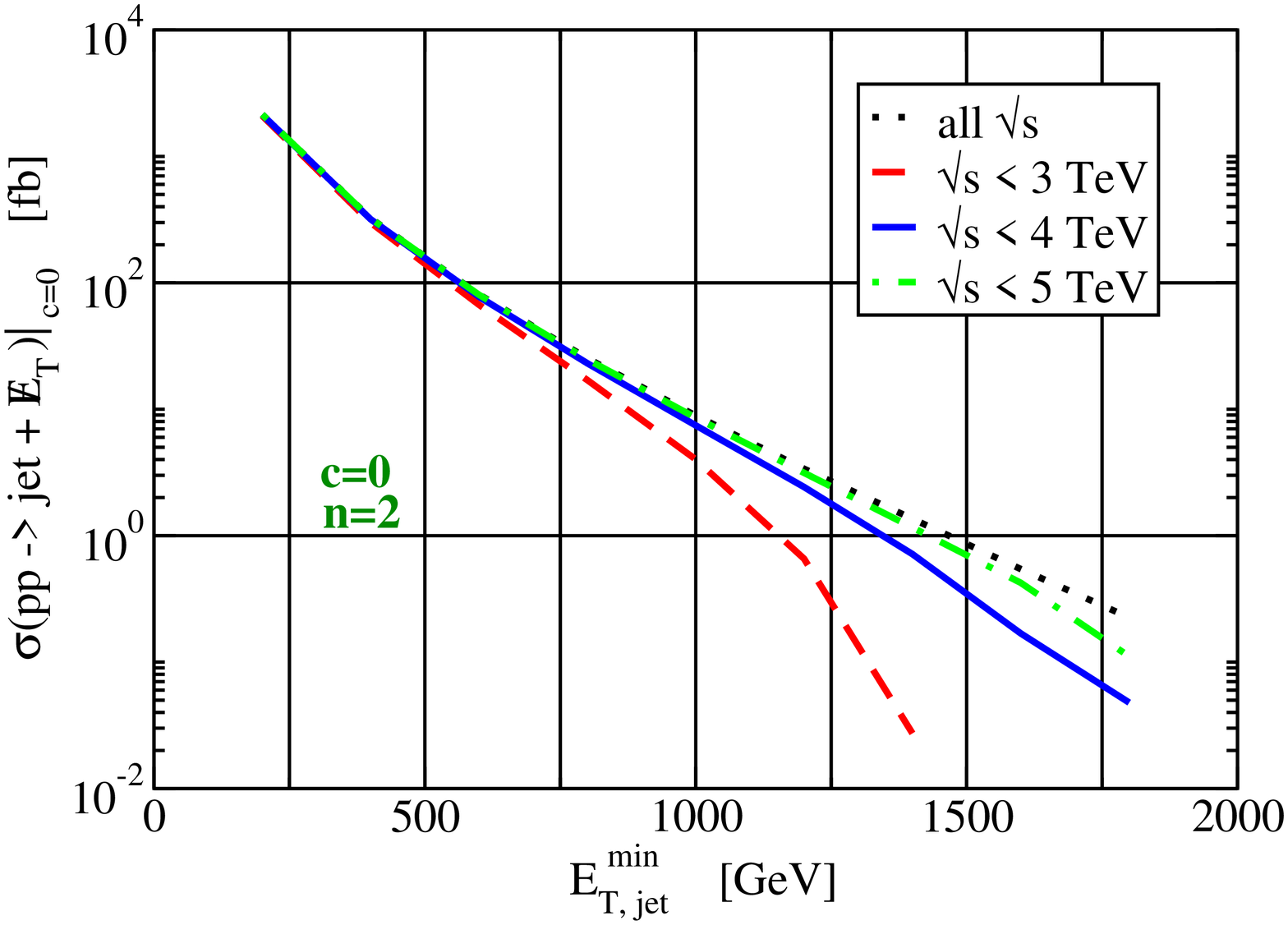, height=6.0cm}}
    \caption{Total jet + nothing cross-section for graviscalar
    production at LHC to evaluate the
       validity limit of the phase space volume when: top) only
       the graviscalar-gluons coupling is
       present; bottom) only the graviscalar-fermion coupling is
       present. The curves are normalized to the cross section for
       graviton production at a value of 
       $E_{T,jet}^{min}$ of 500 GeV.}
   \label{fig:validity}
  \end{center}
\end{figure}

In order to do so, we compare in Fig.~\pref{fig:validity} the
total proton-proton cross-section, $\sigma$, calculated in the
following two ways \cite{realgraviton,Vacavant}:
\begin{enumerate}
  \item We calculate using the above parton cross sections
  for all possible parton energies.
  \item We calculate using the above parton cross sections only
  if $\sqrt{\hat{s}} < M_D$, and set the parton-level cross-section to
  0 if $\sqrt{\hat{s}}> M_D$.
\end{enumerate}
The bottom panel of Fig.~\pref{fig:validity} shows this comparison
when only the quark-graviscalar couplings, $g, g_5$, are nonzero.
The top panel shows the result assuming only the
gluon-graviscalar couplings, $c, b$, do not vanish. The
various curves plot $\sigma$ against $E_{T,jet}^{min} = P_{cut}$
for different choices for $\hat{s}_{max}$ (with the effective dimensionful
couplings $\overline{g}$ and $\overline{c}$ held fixed at some 
arbitrary reference value) When the two curves start to
deviate, high-energy parton contribution are significant and we do
not trust our calculation.

We use these curves to define the maximum value of $P_{cut}$ which
we may trust, given a value for $M_D$ (and so also for the
effective graviscalar couplings). Quantitatively, we fix $P_{cut}$
by demanding that the curves not differ by more than 10\%. As is
clear from the figure, the value of $P_{cut}$ which is obtained in
this way is smaller for the gluon-graviscalar couplings ($g=g_5
=0$) than for the quark-graviscalar couplings ($c =b =0$),
and we use the lower of the two in the following calculations.

Notice that these two plots also indicate that for numerically
equal couplings, the gluon-graviscalar couplings dominate the
cross-section at high energy, while the low-energy regime gets
bigger contributions from the quark-graviscalar interactions. This
is as expected from the effective Lagrangian, since the gluon
terms involve an addition derivative (and so an additional power
of $E/M_D$ in cross sections) relative to the quark terms.

\begin{figure}[htbp]
  \begin{center}
    \mbox{\epsfig{figure=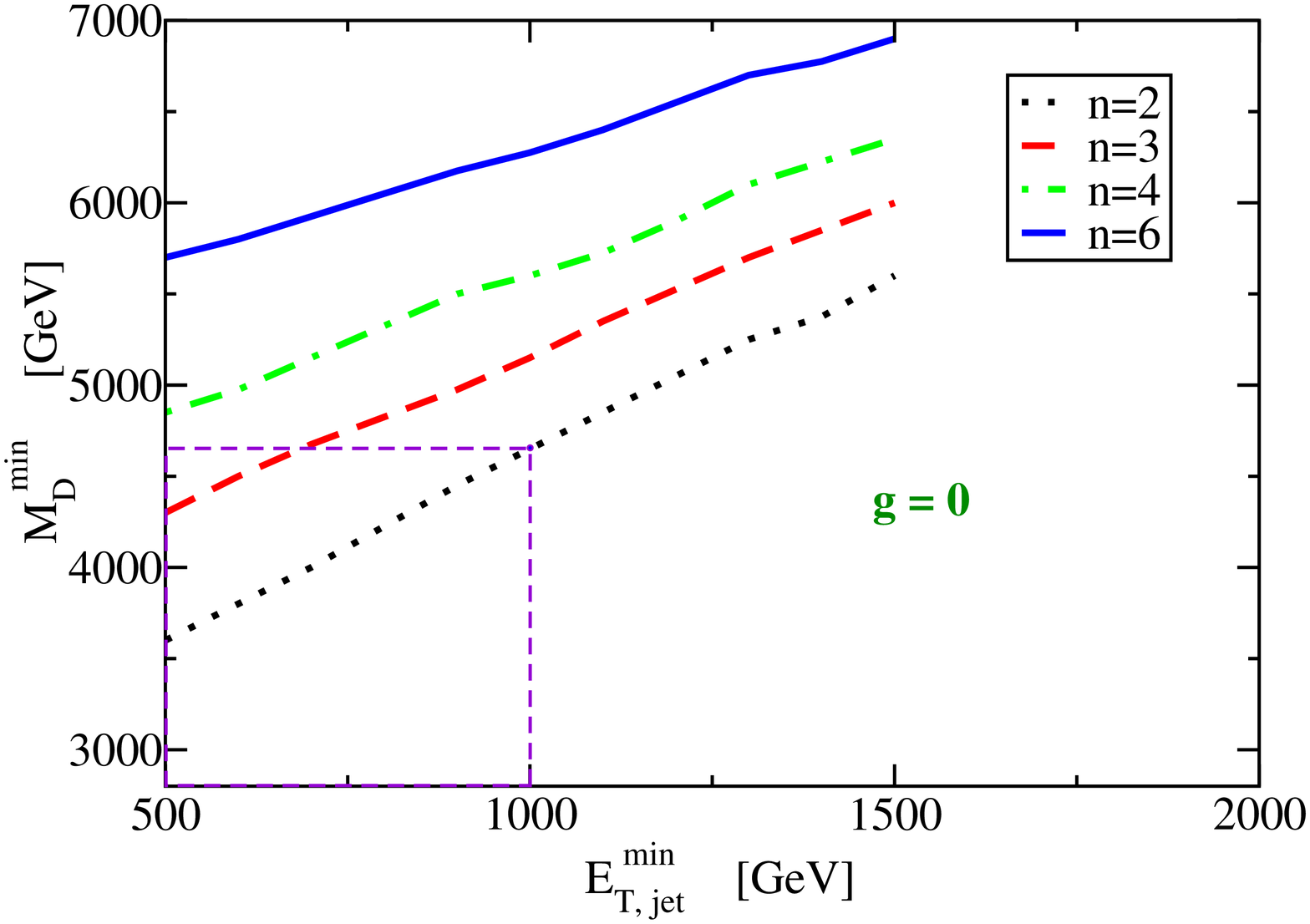, height=5.9cm}}
    \mbox{\epsfig{figure=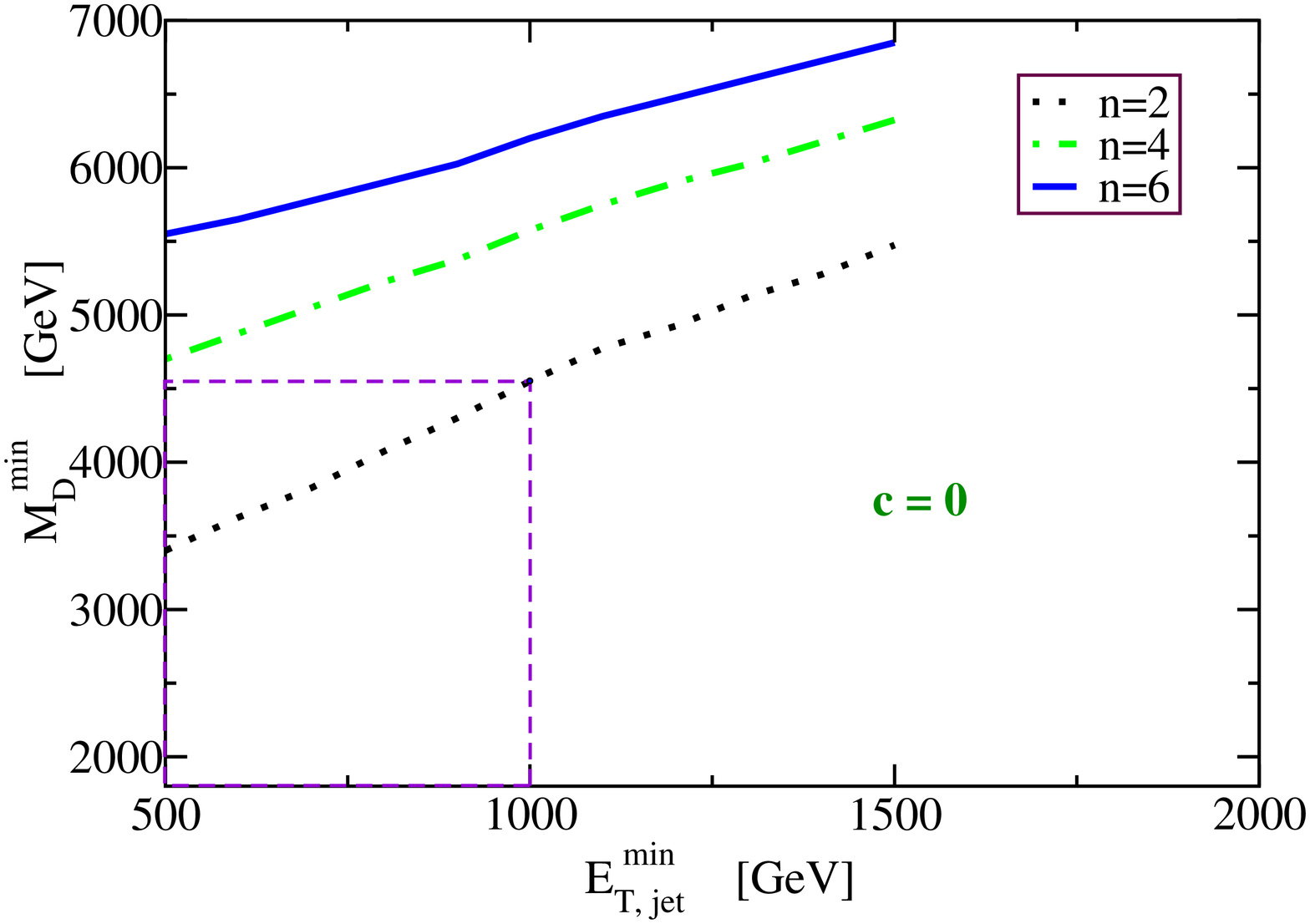, height=5.9cm}}
     \caption{Evaluation of the minimum (4+n)-dimensional Planck scale
     for which the effective model is valid, at the 90\% level, given a cut 
     $E_{T,jet}^{min}$, as explained in the text}
    \label{fig:md}
  \end{center}
\end{figure}

For fixed $P_{cut}$, we may ask how large the effective
graviscalar couplings may be without introducing more than a 10\%
error into our calculations. This is illustrated in
Figs.~\pref{fig:md}, which show the smallest value of $M_D$ which
is permitted given a choice for $E_{T,jet}^{min} = P_{cut}$. This
lower bound on $M_D$ amounts to choosing an upper bound for the
effective dimensional couplings, $\overline{g}$ and 
$\overline{c}$.

In what follows we calculate the proton-proton cross sections
assuming a value for $P_{cut}$, and so the condition $M_D >
M_{D}^{min}$ determines the domain of validity of our
calculations. As discussed above, we determine $M_D^{min}$ using
gluon-graviscalar couplings, since this is the stronger
requirement. From this we find the minimum values for $\overline{g}$ and
$\overline{c}$ for which a graviscalar signal would be observable above the
statistical Standard Model background at the $5\,\sigma$ level.

\section{Simulations}

As mentioned above, the processes discussed in
Sect.~\ref{sect:processes} were implemented using PYTHIA as
external processes. Parton flavors were properly assigned in each
event according to the CTEQ 5L parton distribution functions
evaluated at the renormalization scale $Q^2=\frac12 \, {M^2} +
p_T^2$ and the color flow between those parton was applied. ATLAS
detector effects were incorporated
using the fast Monte Carlo program ATLFAST \cite{ATLFAST}.

For the purpose of comparing with Standard Model backgrounds, we
must choose reference values for the couplings $g$ and $c$ as well
as for the number of extra dimensions $n$. For these purposes a
useful choice is a set of couplings for which the low-energy
approximation applies and for which the proton-proton cross
section is the same size as the cross-section which was determined
as being what is required for a $5\, \sigma$ discovery for
graviton production \cite{Vacavant}. The reference dimensionless effective
couplings, when $P_{cut} = 500$ GeV, $M_D = 5$ TeV and for $n=2$
extra dimensions, are
then 
$g \simeq 0.70$ or $c \simeq 0.41$. These couplings are reasonable
from the point of view of the effective theory because the $g$ and
$c$ we obtain in this way are smaller than unity. Notice that the
existence of such couplings shows that there exist scenarios for
which the graviscalar production would be as important as graviton
production.

We remark that the following analysis is independent of this
choice of reference values.

\subsection{Standard Model Backgrounds}
If graviscalars are produced in association with a jet in proton
collisions the event can be found by searching for a jet plus
missing energy: $pp \to \hbox{jet} + \sla{E}_T$. The Standard
Model background to this process arises from events having
neutrinos in the final state. The principal backgrounds of this
type and their cross-section in the phase space regions
$E_{T,jet}^{min} > 500$ GeV and $E_{T,jet}^{min} > 1000$ GeV are
given, following ref.~\cite{Vacavant}, in Table
\pref{tab:SMbackground}. For comparison, the graviscalar
production cross-section using our reference couplings --- 2 extra
dimensions, $M_D=5$ TeV, $E_{T,jet}^{min}=500$ GeV and
$(c,g)=(0.41,0.70)$ --- is $\sigma = 156$ fb.

\begin{table}[htbp]
  \begin{center}
    \begin{tabular}{|c|c|c|}\hline
      Processes & \multicolumn{2}{c |}{cross-section (fb)} \\ \cline{2-3}
      & 500 GeV & 1000 GeV \\ \hline\hline
      $pp\rightarrow$jet$+Z(\rightarrow\nu\nu)$ & 278 & 6.21\\ \hline
      $pp\rightarrow$jet$+W(\rightarrow e\nu_e)$ & 364 & 8.57\\ \hline
      $pp\rightarrow$jet$+W(\rightarrow\mu\nu_{\mu})$ & 364 & 8.51\\ \hline
      $pp\rightarrow$jet$+W(\rightarrow\tau\nu_{\tau})$ & 363 & 8.50 \\ \hline
    \end{tabular}
    \caption{S.M. background to the graviscalar production at ATLAS
    and their cross-section
    for different phase space volume.}
    \label{tab:SMbackground}
  \end{center}
\end{table}

\begin{figure}[htbp]
  \begin{center}
    \mbox{\epsfig{figure=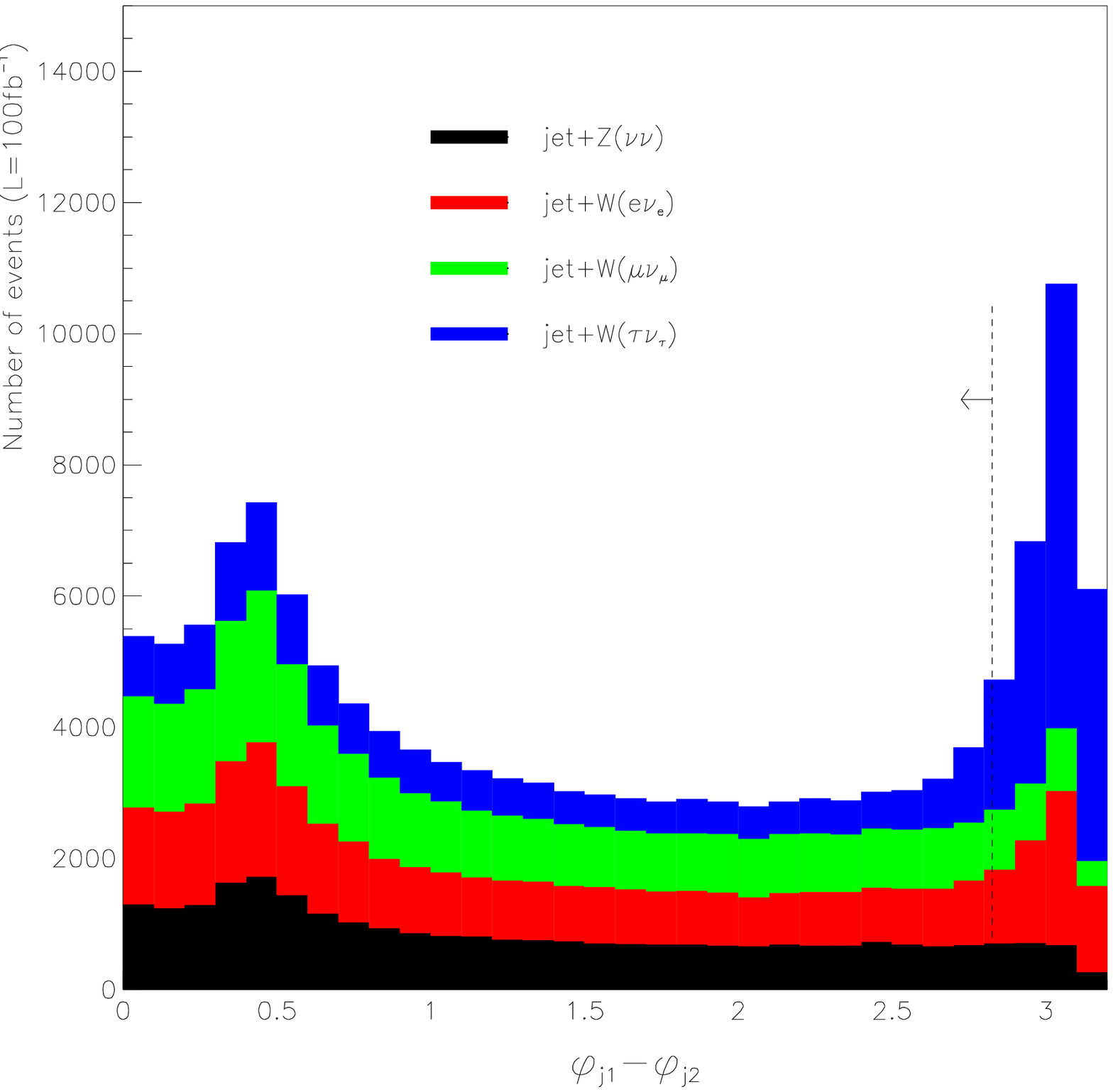, height=7.5cm}}
    \mbox{\epsfig{figure=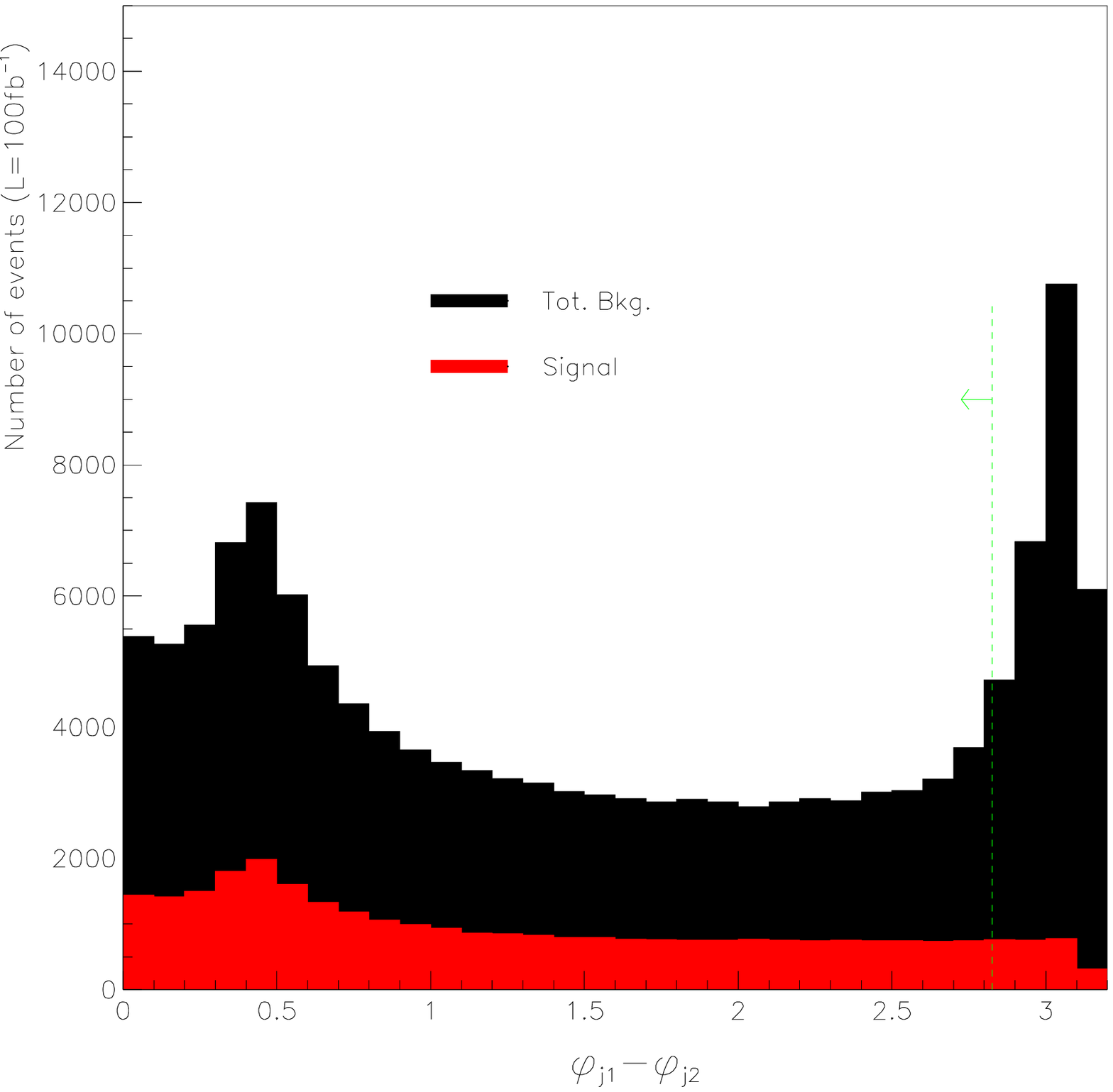, height=7.5cm}}
    \caption{Distribution of the difference in the azimuthal angle between the two most energetic jets
    of an events for: (top) each of the background; (bottom) the signal and the total
    background on top of it.}
    \label{fig:delphi}
  \end{center}
\end{figure}

\subsection{Analysis}
We now determine a set of experimental cuts with which we can find
the minimum values of $g$ and $c$ for which a $5\sigma$ discovery
is possible. 
Jets are reconstructed using the cone algorithm with a cone radius 
$\Delta R = 0.4$.
In the detector ATLAS, leptons are detected
if they are emitted in the range of pseudorapidity $ -2.5 < \eta < 2.5$.
Leptons are defined as isolated if the energy deposited by other particles
in a cone of radius $\Delta R = \sqrt{(\Delta\phi)^2 + (\Delta\eta)^2}$
is less that 10 GeV.
We impose the following two cuts.

As a first cut we require:
\begin{quote}
{\bf Cut 1:} No isolated lepton (electron or muon) with $p_T > 6$ GeV
is allowed in the event. 

\end{quote}
This eliminates most of the $W \to e\nu_{e}$ and $W \to
\mu\nu_{\mu}$ events, leaving only those for which the leptons are
not properly reconstructed in the detector. It does not eliminate
events like $W \to \tau \nu_\tau$ in which the $\tau$ decays
hadronically, however in this case we expect to also have an
energetic low-multiplicity jet which is opposite, in the azimuthal
plane, to the principal jet. 
We see in Fig.~\pref{fig:delphi} that
a cut on the difference in azimuthal angle between the two most
energetic jets can eliminate a significant fraction of this $\tau$
background.

We therefore impose the second cut:
\begin{quote}
{\bf Cut 2:} We keep only events for which
$|\varphi_{j_1}-\varphi_{j_2}| < 2.285$ radians.
\end{quote}
This cut is chosen to maximize the significance of the remaining
signal.

\begin{figure}
  \begin{center}
    \mbox{\epsfig{figure=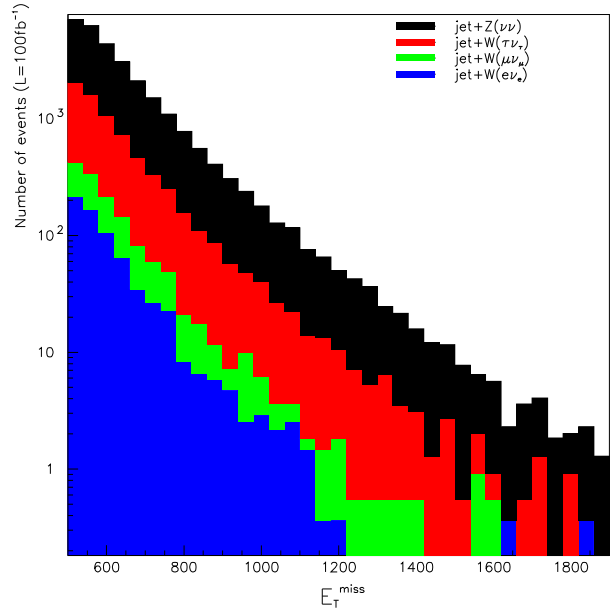, height=9.0cm}}
    \caption{Contributions of different processes to the total background
     after application of the cuts, for an integrated luminosity of 100 fb$^{-1}$.}
    \label{fig:ptdist}
  \end{center}
\end{figure}

Fig.~\pref{fig:ptdist} shows the relative contribution of each
process to the total background. After the above cuts, the most
important background is $pp \to \hbox{jet} Z \to \hbox{jet}~
\nu\nu$. Table~\pref{tab:cuts} makes this more explicit, by
breaking down the background and comparing it and the signal
before and after the cuts are applied, assuming an integrated
luminosity of 100 fb$^{-1}$. For this integrated luminosity
--- which corresponds to one year's running at the nominal LHC
luminosity --- we therefore expect a total of 36700 background
events to remain after cuts.

\begin{table}[htbp]
  \begin{center}
    \begin{tabular}{|c|c|c|c|}\hline
      Processes  & $E_T^{min}>500$GeV & \# events after cut 1 & \#
      events after cut 2\\ \hline\hline
      jet$+Z(\rightarrow\nu\nu)$ & 27760 & 27100 & 24940\\ \hline
      jet$+W(\rightarrow e\nu_e)$ & 36420 & 5224 & 1430\\ \hline
      jet$+W(\rightarrow\mu\nu_{\mu})$ & 36370 & 957 & 866\\ \hline
      jet$+W(\rightarrow\tau\nu_{\tau})$ & 36330 & 24600 & 9459\\ \hline
      jet+Graviscalar & 30960 & 30090 & 27720\\ \hline
    \end{tabular}
    \caption{Number of signal ($M_D=5$ TeV, $n=2$, $g=0.70$ and $c=0.41$) and
      background events that survive each cut for an integrated luminosity of
      100 fb$^{-1}$. The cuts are defined in the text.}
    \label{tab:cuts}
  \end{center}
\end{table}

Is this good enough? We estimate the number of signal events
required for a $5\sigma$ discovery using the following
significance criterion:
\begin{equation}
  \frac{S}{\sqrt{S+B}}>5 \,,
\end{equation}
where $S$ and $B$ are respectively the number of signal and
background events. For this many background events we therefore
have a $5\sigma$ discovery if more than 970 graviscalar events are
detected, {\it i.e.} if the total cross-section for the process
is: $\sigma(pp \to \hbox{jet} + \phi) > 10.9$ fb. For $n=2$ this
corresponds to the effective couplings:
\begin{equation}
\overline{g} > 0.18 \mathrm{~TeV}^{-1} \mathrm{ ~~~~if~~~~  }
\overline{c}=0 \,,
\end{equation}
or
\begin{equation}
\overline{c} > 3.2\times10^{-1} \mathrm{~TeV}^{-2} \mathrm{ ~~~~if~~~~  }
\overline{g}=0 \, .
\end{equation}

\begin{figure}[h]
  \begin{center}
    \mbox{\epsfig{figure=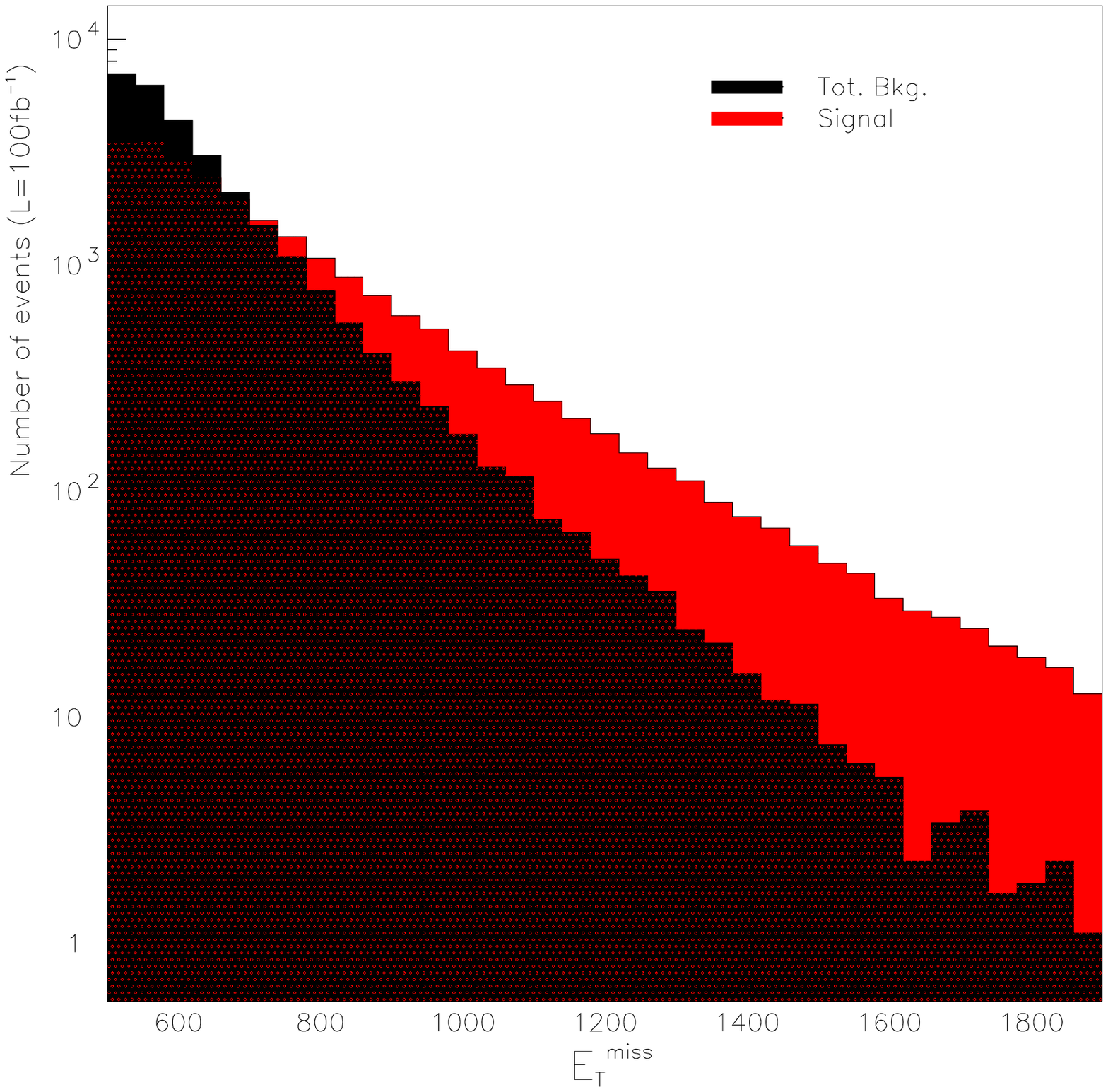, height=7.8cm}}
    \mbox{\epsfig{figure=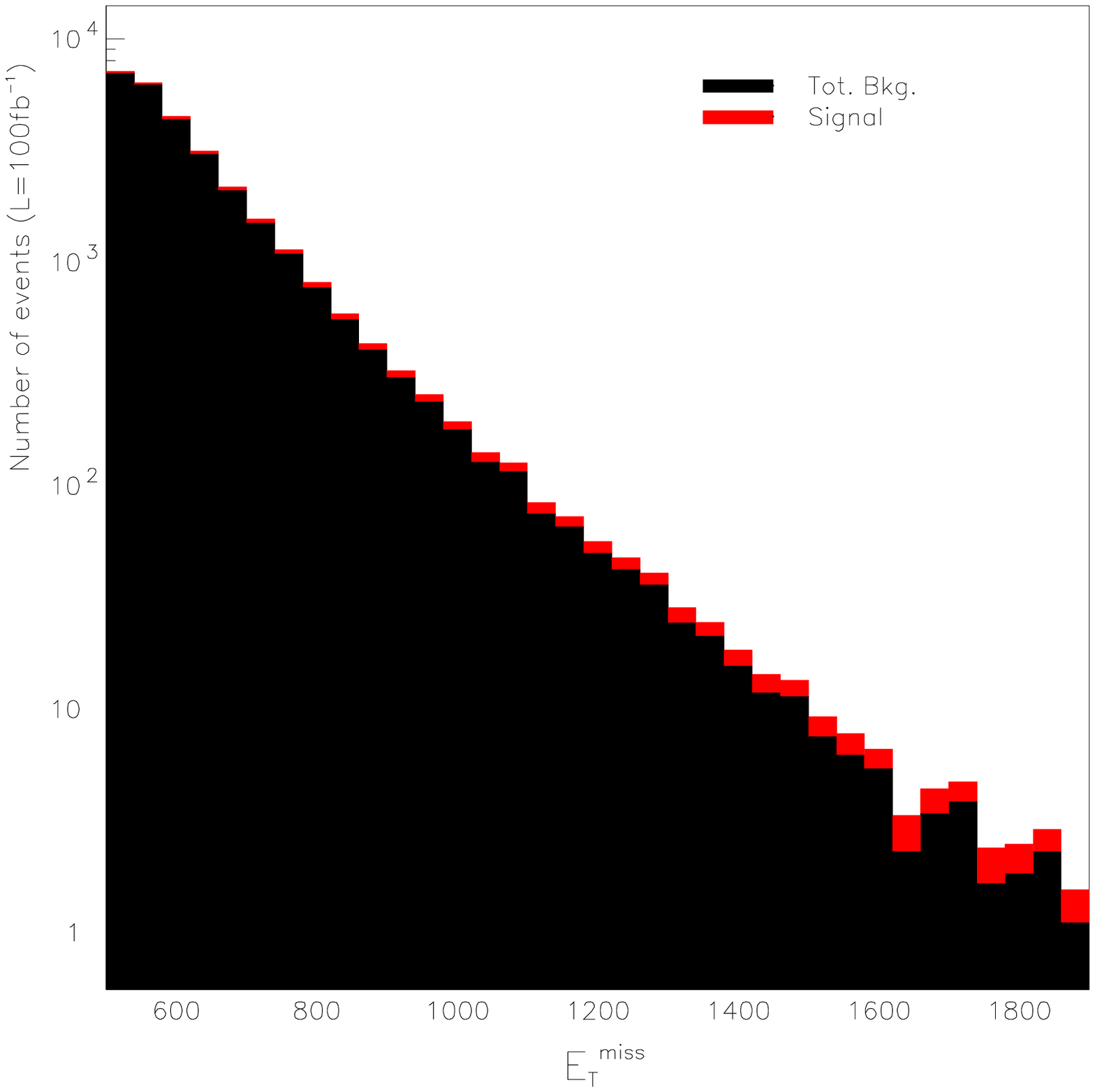, height=7.8cm}}
    \caption{Distribution of missing $E_T$ for background and
    signal when: (top) the cross-section
             for graviscalar production is the same as the
             graviton one; (bottom) this cross-section
             is at its discovery limit.}
    \label{fig:etmiss}
  \end{center}
\end{figure}

Fig.~\pref{fig:etmiss} plots the cross section as a function of
missing energy, for two choices of couplings. The choice in the
second panel corresponds to the $5\sigma$ discovery limit, and
shows that the discovery would be due to an excess of events in
the distribution of missing transverse energy at high energies.

Notice that the cross section required for discovery is determined
by the background rate and so is the same for all choices for $n$,
the number of extra-dimensions. In fact, as can be seen in
eq.~\pref{equ:cross-sect}, the $\hat{t}$ and $\hat{u}$ dependence
of the cross-section does not depend on the number of extra
dimensions. Only the effective coupling constants and the energy
dependence depend on $n$, through the overall factor
$(M^2)^{({n-2})/{2}}$. Since the angular distribution depends only
on $\hat{t}$ and $\hat{u}$, Cut 2 has the same effect for all
possible $n$ (as does Cut 1).

\subsection{Results}
We now turn to the central question: What range of effective
couplings are likely to be detectable at the LHC?

We have seen that any determination of the reach of LHC must be
made relative to a choice for $E_{T,jet}^{min} = P_{cut}$, since
this plays a role in the reliability of the entire theoretical
calculation. From Fig.~\pref{fig:md} we see that the choice
$E_{T,jet}^{min} = 500$ GeV implies that the cross-section is
sensitive to high-energy parton processes at less than the 10\%
level, provided $M_D \ge M_D^{min}$, where $M_D^{min} =3.60$,
$4.30$, $4.85$ and $5.70$ TeV for $n = 2$, 3, 4 and 6 extra
dimensions respectively. Given the value for $M_D^{min}$ we then
determine what values of couplings produce an observably large
cross section ({\it i.e.} $\sigma > 10.9$ fb).

Suppose $(\overline{g}_{obs}^{(n)},\overline{c}_{obs}^{(n)})$ are a pair of
dimensionful couplings which each by itself produces a $5\sigma$ signal 
(in n extra-dimensions) above the
Standard Model background. The LHC then can detect couplings which
lie in the intervals
\begin{eqnarray*}
 1 \gtrsim g > \overline{g}_{obs}^{(n)} (\overline{M}_D^{min})^{{n}/{2}}
  \mathrm{ ~~~~if~~~~  } c=0\\
\mathrm{or~~~~~~~~~~~~~~~~~~~~  }\\
 1 \gtrsim c > \overline{c}_{obs}^{(n)} (\overline{M}_D^{min})^{1+n/2}
 \mathrm{ ~~~~if~~~~  } g=0 \,.\\
\end{eqnarray*}
The upper limit in these inequalities expresses the theoretical
criterion that the calculation only makes sense below the cut-off
scale $M_D^{min}$.

\begin{figure}
  \begin{center}
    \mbox{\epsfig{figure=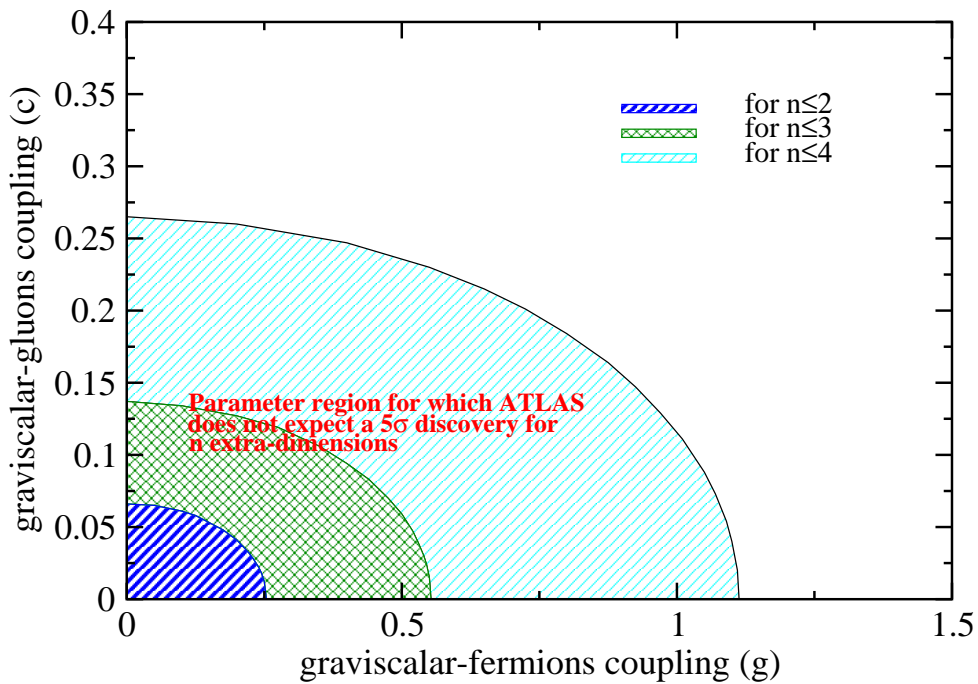, height=9.0cm}}
    \caption{Parameter region for graviscalar theory at
    second order in $E/M_D$ that allow testable and valid
      physical prediction at ATLAS, for different number
      of extra-dimension. The graviscalar-fermion and graviscalar-gluon
dimensionless couplings are effectively combinations 
$(g^2+g_5^2)^{1/2}$ and $(c^2+c_5^2)^{1/2}$}
    \label{fig:cg}
  \end{center}
\end{figure}

Fig.~\pref{fig:cg} shows the couplings which are accessible if
both $c$ and $g$ are simultaneously nonzero. The shaded regions
indicate the range of couplings which are too small to have
detectable effects for various choices for the dimension $n$. The
potentially observationally-interesting couplings are those which
lie outside the ellipses, but inside the box defined by $c < 1$
and $g < 1$. (Recall that since only $c^2$ and $g^2$ enter into
the cross sections, these plots should be interpreted as
constraints on $|c|$ and $|g|$.)

We see that whether useful constraints are possible depends on the
number of dimensions $n$, and fewer dimensions gives better reach.
If $n = 2$ couplings outside the innermost region are potentially
detectable. For $n = 3$ detectable couplings must lie outside the
next-to-innermost region.  Detection for the case n=4 is unlikely
for the dimensionless constant $g$ in the limit of small $c$, but
is possible if $c$ is also nonzero. For more then 6 extra
dimensions, detection of any signal is unlikely since the minimum
value of couplings needed for a discovery when $n=6$ is $(c,g) =
(0.9, 3.3)$. Any coupling in this region is far enough from the
limits of validity of the calculation to have confidence in the
result.

\begin{table}[htbp]
  \begin{center}
    \begin{tabular}{|c|c|c|}\hline
      ndim   & $M_D^{min}$ (TeV) & $M_D^{max}$ (TeV)\\ \hline\hline
      2 & 3.2 & 14.00  \\ \hline
      3 & 3.8 & 6.45  \\ \hline
      4 & 4.4 & 1.45   \\ \hline
    \end{tabular}
    \caption{Sensitivity of the ATLAS detector to the fundamental Planck scale $M_D$
         through the discovery of a graviscalar signal, for $c = 0$ and for an integrated
         luminosity of 100 fb$^{-1}$. For $n \ge 4$, observation of a signal is not possible.}
    \label{tab:result}
  \end{center}
\end{table}

An alternative way of expressing the potential ATLAS reach for a
graviscalar signal is in terms of the value of the fundamental
Planck scale to which the detector might be sensitive. It is
bounded on the low side by the requirement that the effective
theory be a good approximation, and on the high side by the
condition that the signal be detectable. Defining the effective
upper limit by
\begin{equation}
  M_D^{max} = (2\pi)^{\frac{n}{n+2}} \times
  \min \left[ \left( \overline{g}_{obs}^{(n)} \right)^{-2/n},
  \left(\overline{c}_{obs}^{(n)} \right)^{-2/(n+2)} \right] \,,
\end{equation}
there are prospects for detection when $M_D^{min} \lsim
M_D^{max}$. Table ~\pref{tab:result} summarizes these results,
with $M_D^{max}$ calculated using the worst case: $c=0$.

In the more optimistic limit $g=0$, the maximum value of the
fundamental Planck mass to which the ATLAS detector is sensitive
increases from 6.45 to 9.50 TeV when $n=3$, and from 1.45 to 7.55
when $n=4$. These results are summarized in Table
\pref{tab:result_g}. We again find that $n=6$ is the limiting case
since for $g=0$, we have $M_D^{max} = 5.8$ TeV $\approx
M_D^{min}=5.7$ TeV.

\begin{table}[htbp]
  \begin{center}
    \begin{tabular}{|c|c|c|}\hline
      ndim   & $M_D^{min}$ (TeV) & $M_D^{max}$ (TeV)\\ \hline\hline
      2 & 3.60 & 14.10  \\ \hline
      3 & 4.30 & 9.50  \\ \hline
      4 & 4.85 & 7.55   \\ \hline
      6 & 5.70 & 5.80   \\ \hline
    \end{tabular}
    \caption{Sensitivity of the ATLAS detector to the fundamental Planck scale $M_D$
         through the discovery of a graviscalar signal, for $g = 0$ and for an integrated
         luminosity of 100 fb$^{-1}$. For $n \le 6$, observation of a signal is possible.}
    \label{tab:result_g}
  \end{center}
\end{table}

A comparison with the results obtained from graviton emission
\cite{Vacavant} is also instructive, although some care must be
taken in so doing because the graviton results were obtained using
a more restrictive phase-space cut ($E_T^{min}>1$ TeV), a
different criterion for defining the validity region of the model
and with a more conservative statistical estimator (${S}>
{\sqrt{7B}}$). Tables \pref{tab:result2c} and \pref{tab:result2g}
compare the sensitivity of ATLAS to $M_D$ as computed using
graviscalar and graviton production, using these more conservative
criteria. The two tables differ in their choice of either $g = 0$
or $c = 0$.

Table \pref{tab:result2c} also shows the existing non-accelerator
limit on $M_D$, taken from ref.~\cite{March-Russell} (see also
\cite{SusyLED}). Unlike the situation for gravitons (which couple
universally) these astrophysical bounds are more model-dependent
when applied to graviscalars. This model dependence arises because
they directly bound the couplings of KK modes to electrons and
photons, and so need not directly apply to the gluon and quark
couplings of most interest for colliders.

\begin{table}[htbp]
  \begin{center}
    \begin{tabular}{|c|c|c|c|c|c|c|}\hline
      c=0  & \multicolumn{2}{c |}{Graviton} & \multicolumn{2}{c |}{Graviscalar}
      & \multicolumn{2}{c |}{limit from cosmology}\\ \hline
      & $M_D^{min}$ & $M_D^{max}$ & $M_D^{min}$ & $M_D^{max}$  & $M_D^{min}$ (A)
      & $M_D^{min}$ (B)\\ \hline\hline
      n=2 & $\sim$4.0 TeV & 7.5 TeV & 4.35 TeV & 5.45 TeV & O(90) TeV & $\sim 10$ TeV \\ \hline
      n=3 & $\sim$4.5 & 5.9 & 4.85 & 3.65  & 5.0 & 0.8\\ \hline
      n=4 & $\sim$5.0 & 5.3 & 5.35 & 3.20  & $\lesssim4$ & $\lesssim1$  \\ \hline
    \end{tabular}
    \caption{With c=0: comparison of the sensitivity of ATLAS to $M_D$ for graviscalar
       and graviton signals under the conditions $E_T^{min}>1$TeV, $\frac{S}{\sqrt{7B}}$ and
       with indirect constaints from cosmology. The integrated luminosity
       is 100 fb$^{-1}$. For $n \ge 3$, observation of a graviscalar signal is not possible since
       $M_D^{min}>M_D^{max}$. For the cosmology bounds, Scenario A means limits to neutron star
       heating by KK-decays, while scenario B corresponds to bounds from the cooling of SN1987A
       by KK-mode emission.}
    \label{tab:result2c}
  \end{center}
\end{table}

\begin{table}[htbp]
  \begin{center}
    \begin{tabular}{|c|c|c|c|c|}\hline
      g=0  & \multicolumn{2}{c |}{Graviton} & \multicolumn{2}{c |}{Graviscalar} \\ \hline
      & $M_D^{min}$ & $M_D^{max}$ & $M_D^{min}$ & $M_D^{max}$  \\ \hline\hline
      n=2 & $\sim$4.0 TeV & 7.5 TeV & 4.65 TeV & 10.20 TeV \\ \hline
      n=3 & $\sim$4.5 & 5.9 & 5.15 & 7.75 \\ \hline
      n=4 & $\sim$5.0 & 5.3 & 5.60 & 6.50 \\ \hline
    \end{tabular}
    \caption{With g=0: comparison of the sensitivity of ATLAS to $M_D$ for graviscalar
       and graviton signals under the consitions $E_T^{min}>1$TeV, $\frac{S}{\sqrt{7B}}$ and
       with indirect constaints from cosmology. The integrated luminosity
       is 100 fb$^{-1}$. For $n \le 4$, observation of a graviscalar signal is possible.}
    \label{tab:result2g}
  \end{center}
\end{table}

We see again that the scenario where $g \to 0$ gives the best case
for detection, and this is competitive with the graviton result.
We also see that although accelerator experiments are most
sensitive to lower $n$, for quark couplings these may be
pre-empted by the non-accelerator bounds.

The difference between the cases $c=0$ and $g=0$ show that ATLAS
is likely to be only weakly sensitive to the graviscalar Yukawa
couplings (especially keeping in mind these are naturally expected
to be at most of order $v/\overline{M}_D$, as explained in section
2), and a discovery is more likely to come from
gluon-graviscalar couplings. However, once a signal is seen we are
unlikely to be able to decide directly on the relative importance
between $g$ and $c$. Therefore, even if the discovery of a
si\-gni\-fi\-cant graviscalar signal at ATLAS should turn out to be
possible, it is unlikely to completely fix its couplings.

\subsection{Graviton-Graviscalar Confusion}
Should a missing-energy signal be seen at the LHC, how does one
tell if it is due to gravitons or graviscalars? We do not yet see
a way to do so, despite the difference in their spin, for the
following reasons.

\begin{itemize}
\item The graviscalar production cross-section has an energy dependence
which is similar to the graviton one, precluding the use of
$P_{T,jet}$, $\sla{E}_T$ or any other function of energy to
discriminate the two.
\item Parton-level discriminants are not likely to be of practical
use, because the center of mass energy of the hard scattering is
not known in a $pp$ collider such as the LHC. Furthermore, the
final state we consider consists of a single jet and missing
transverse energy, so it is not possible to reconstruct the
longitudinal component of momentum of the system of interacting
partons, nor their angular distribution in the center of mass, nor
their forward-backward asymmetry. Even if this were possible, we
have checked that the discrimination between the shapes of the
graviscalar and graviton differential cross sections is difficult
even at the purely theoretical parton level. Only gluon fusion
processes lead to a small difference.
\end{itemize}

\section{Conclusions}
We have computed the rate for the production of extra-dimensional
scalars (as opposed to components of extra-dimensional tensors
which look at low energies like 4D scalars) in $pp$ collisions.
Such particles arise in virtually all supersymmetric
higher-dimensional theories, and our work complements previous
studies of gravitons \cite{realgraviton, Vacavant} and of extra-dimensional
vectors \cite{SusyLED}. Because of the way we compute our phase
space integrals, our study applies to large-extra-dimensional
(ADD-type) models and not to warped (RS-type) models.

We find that the cross sections for the reaction $pp \to \phi +
\hbox{jet}$ are similar in size and shape to those for graviton
production, although the competing non-accelerator constraints on
the couplings can differ because graviscalars need not couple
universally (unlike gravitons). We used simulation codes tailored
to the ATLAS detector, and conclude that ATLAS can be sensitive to
graviscalar couplings, provided there are less than 6 extra
dimensions. The sensitivity improves with fewer dimensions,
although so does the restrictiveness of non-accelerator bounds on
the extra-dimensional Planck scale. Nontrivial windows of
opportunity can be consistent with all bounds.

Generically, $pp$ collisions are more sensitive to graviscalar
couplings to gluons than they are to couplings to quarks. Both
couplings are turned on in our analysis, and we find that
observable quark couplings often push the limits of validity of
the effective-field-theory description.

\section{Acknowledgements}
This work has been performed within the ATLAS
collaboration. We
have made use of physics analysis and simulation tools which are the result of
collaboration-wide efforts.
We would like to acknowledge partial funding from NSERC (Canada).
C.B. research is partially funded by FCAR (Qu\'ebec) and McGill
University.

\newpage

\section{Appendix}
To proceed the cross sections, we require the Feynman rules for the $\bar{q}q\phi$,
$gg\phi$ and $ggg\phi$ vertices which follow from the effective
lagrangian of eq.~(\ref{E:efflagrangian}), which are:

\vskip 0.5in
\epsfxsize=1.5in \epsfbox{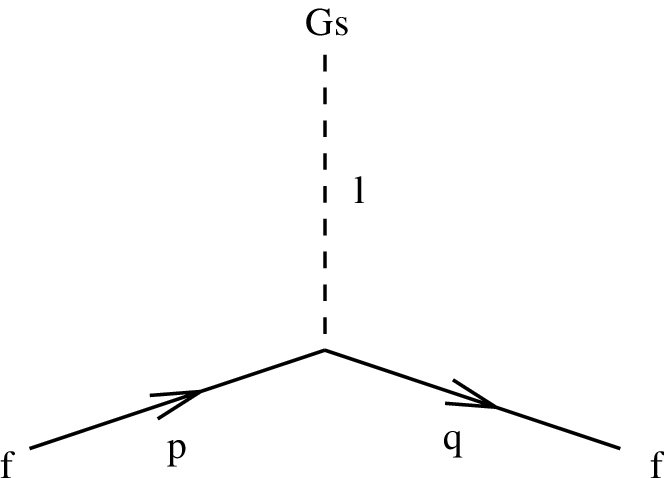}  \hspace{0.5 in} $= -i(\overline{g}+i\overline{g}_5\gamma_5)$

\vskip 0.5in
\epsfxsize=1.5in \epsfbox{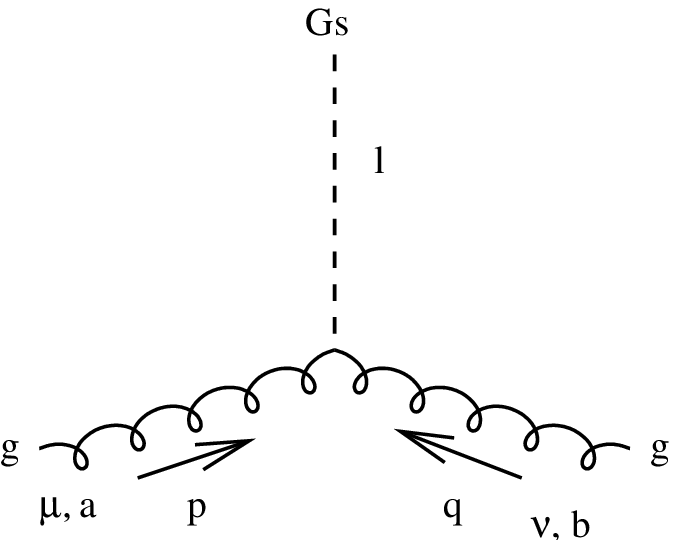}
\hspace{0.5 in}
$=4i[\overline{c}(p.q)g_{\mu\nu} - \overline{c}p_{\mu}q_{\nu} -
\overline{b}\epsilon_{\mu\nu\alpha\beta}p^{\alpha}q^{\beta}]\delta_{ab}$

\vskip 1in
\epsfxsize=1.5in \epsfbox{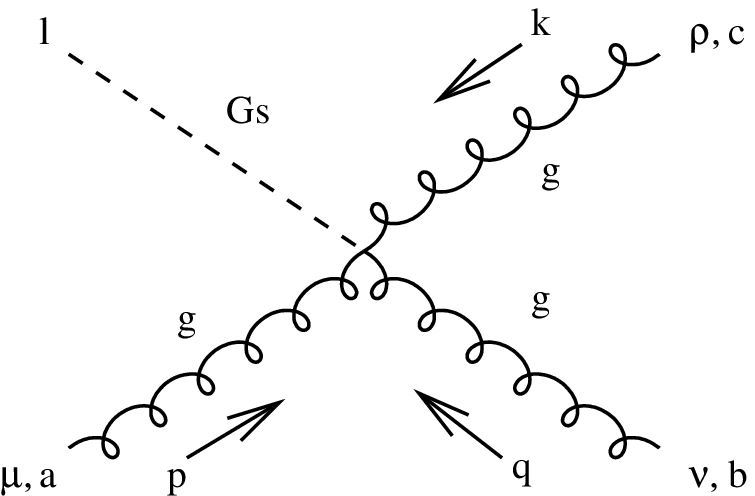}
\begin{center}
$=4g_3f^{abc}[\overline{c}g_{\mu\nu}(p_{\rho}-q_{\rho}) + \overline{c}g_{\mu\rho}(k_{\nu}-p_{\nu})
+\overline{c}g_{\nu\rho}(q_{\mu}-k_{\mu}) - \overline{b}\epsilon_{\alpha\mu\nu\rho}(p^{\alpha}
«+q^{\alpha} + k ^{\alpha})]$
\end{center}

\begin{figure}[htbp]
  \begin{center}
    \epsfxsize=5.0in \epsfbox{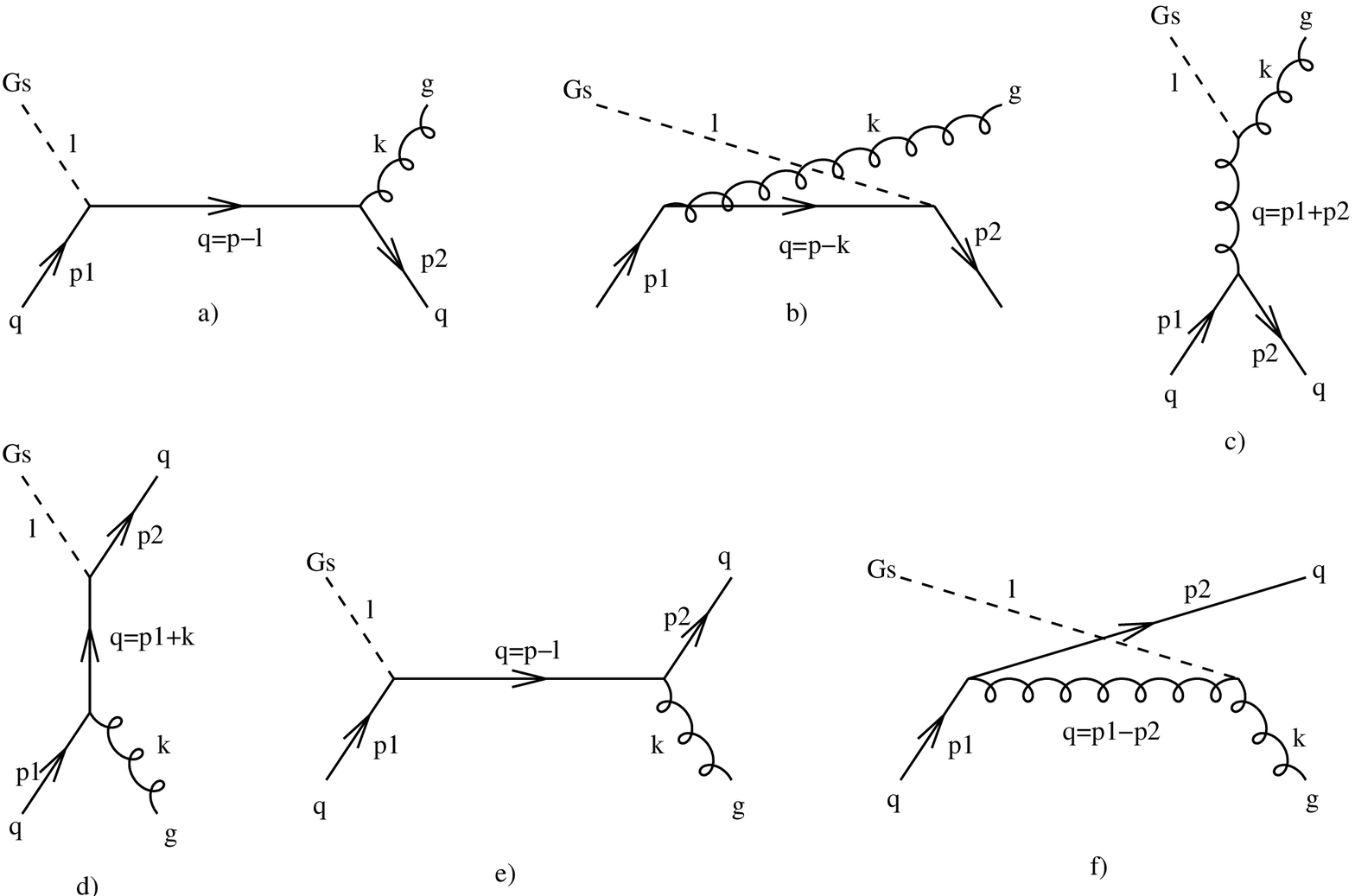}
    \epsfig{file=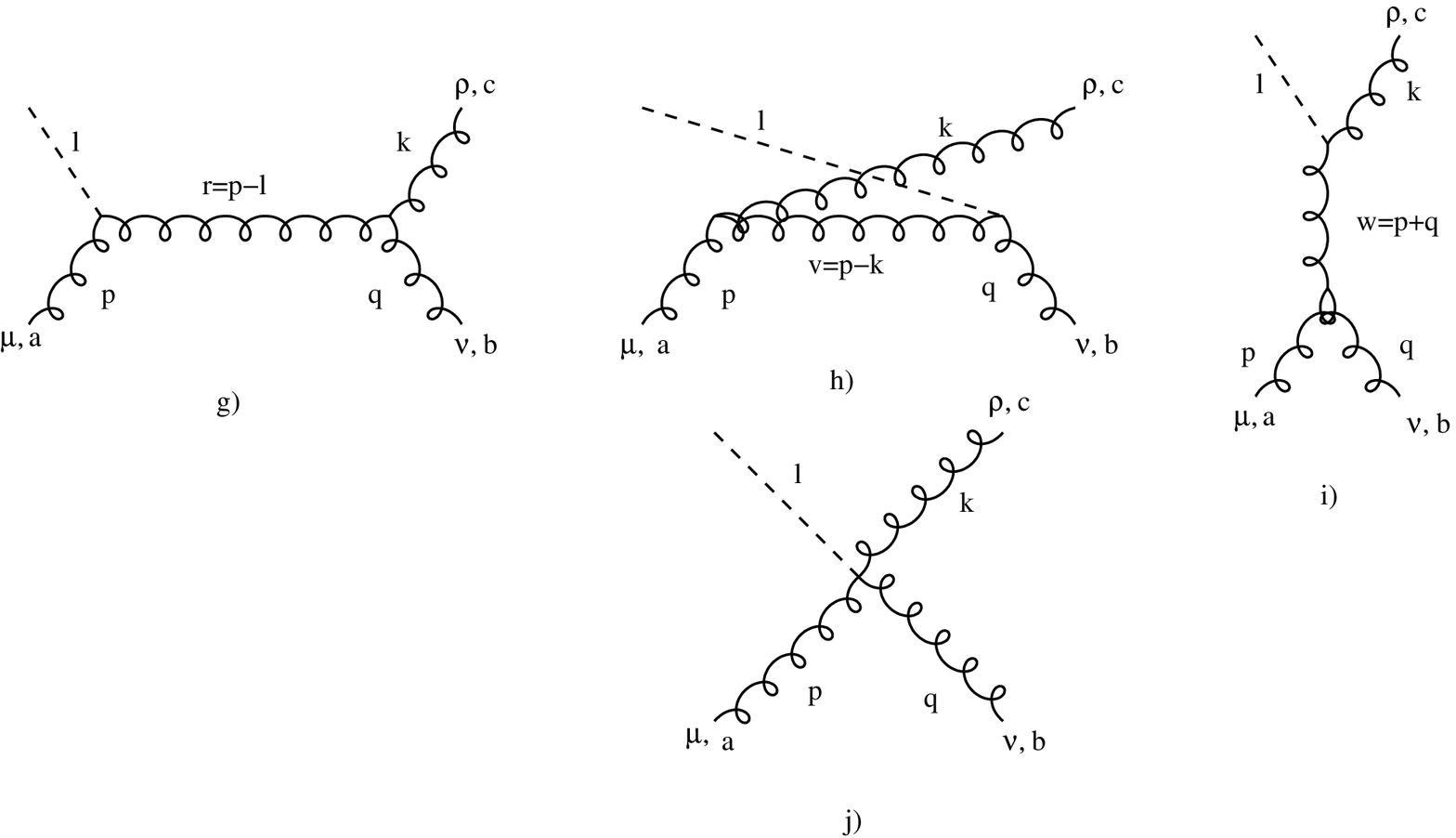, width=5.0in}
   \caption{The parton-level Feynman graphs which contribute to graviscalar
   production with an associated jet in proton-proton scattering.}
    \label{fig:feyngraph}
  \end{center}
\end{figure}

\end{document}